\documentclass[twocolumn,showpacs,preprintnumbers,amsmath,amssymb]{revtex4-1}
\usepackage{graphicx}
\usepackage{textcomp}

\begin{document}
\title{Coherent perfect absorbers for transient, periodic or chaotic optical fields: time-reversed lasers beyond threshold}
  \normalsize
\author{Stefano Longhi 
and Giuseppe Della Valle}
\address{Dipartimento di Fisica, Politecnico di Milano and Istituto di Fotonica a Nanotecnologie del Consiglio Nazionale delle Ricerche, Piazza L. da Vinci
32, I-20133 Milano, Italy}

%
\bigskip
\begin{abstract}
\noindent Recent works [Y.D. Chong {\it et al.}, Phys. Rev. Lett. {\bf 105}, 053901 (2010);
W. Wan {\it et al.}, Science {\bf 331}, 889 (2011)] have shown  that the time-reversed process of lasing at threshold realizes a coherent perfect absorber (CPA). In a CPA, a lossy medium in an optical cavity with a specific degree of dissipation, equal in modulus to the gain of the lasing medium,  can perfectly absorb coherent optical waves at discrete frequencies that are the time-reversed counterpart of the lasing modes. Here the concepts of time-reversal of lasing and CPA are extended for optical radiation emitted by a laser operated in an arbitrary regime, i.e. for transient, chaotic or periodic  coherent optical fields. We prove that any electromagnetic signal $E(t)$ generated by a laser system \textbf{S} operated in an arbitrary regime can be perfectly absorbed by a CPA device $\bf{S'}$  which is simply realized by placing inside \textbf{S} a broadband linear absorber (attenuator) of appropriate transmittance. As examples, we discuss CPA devices that perfectly absorb a chaotic laser signal and a frequency-modulated optical wave. 
\end{abstract}

\pacs{42.55.Ah, 42.65.Sf}


\maketitle

\section{Introduction}
 A laser oscillator  is a device that self-organizes to emit a narrow-band coherent electromagnetic radiation when the pumping level exceeds a threshold value. 
 Above the first lasing threshold, lasers are generally complex non-linear systems  \cite{L1} and may emit different kinds of output signals $E(t)$, ranging from the ideal monochromatic wave under single-mode continuous-wave operation, to irregular or chaotic signals when multi-mode or other kinds of instabilities set in \cite{L1,L2}, or giant pulses or periodic trains of ultrashort optical pulses when operated in the Q-switching or mode-locking regimes \cite{L3,L4}. Generally speaking, a "time-reversed" laser refers to a device that, rather than emitting the signal $E(t)$ propagating outgoing from the laser cavity, it is capable of perfectly absorbing the same  signal  $E(t)$ that propagates backward into the cavity, without any reflection \cite{Stone1}.  Such a device thus realizes a coherent perfect absorber (CPA) for the signal $E(t)$. For a continuos-wave field, it is known 
that a dissipative resonator can perfectly absorb incident light at the condition of critical coupling \cite{CC}. Recently, Douglas Stone and coworkers have suggested and proven rather generally the possibility to realize a CPA exploiting the time-reversed process of lasing {\it at threshold}.  In this case, the laser system behaves as a linear one, and the time-reversal process is realized simply by the replacement $\epsilon(\mathbf{r}) \rightarrow  \epsilon^*(\mathbf{r})$ for the complex dielectric constant  $\epsilon$ of the medium. In this way, any {\it linear} Êabsorbing medium of {\it arbitrary} shape behaves as a CPA at some discrete frequencies under appropriate coherent illumination (the time reverse
of the output  lasing modes) and provided that a precise amount of dissipation in the medium occurs  (equal in modulus to the threshold gains for lasing).  The CPA idea has received a great interest and stimulated theoretical and experimental studies along different lines \cite{LonghiPRA2010,StoneScience2011,StonePRL2011,uffa,uffa2,uffa3,uffa3bis,StoneP,uffa4,uffa5}.  In particular, in Ref.\cite{StoneScience2011} an experimental demonstration of interferometric control of the absorption based on CPA was reported using a thin slice of silicon illuminated by two beams, whereas in Refs.\cite{LonghiPRA2010,StonePRL2011} a laser-absorber device, which can operate as a CPA and as a laser simultaneously, has been suggested combining the CPA and $\mathcal{PT}$-symmetry concepts. Time-reversal of other optical instabilities, such as time-reversal of optical parametric oscillation, have been proposed to realize a multicolor CPA in Ref.\cite{uffa2}. Also, extension of the concept of time-reversed laser and CPA to the spaser and plasmonic nanostructures have been suggested in Refs.\cite{uffa3bis,StoneP}. Here the energy of the incoming wave is fully transferred into surface plasmon oscillations and evanescent electromagnetic
fields. CPA-based devices may have potential applications to the realization of a new class of absorptive interferometers and nanosensors. In all previous studies, time-reversal of lasing has been limited to consider either a laser at threshold or above threshold in steady-state operation. However, as previously mentioned, a laser can operate in rather complex or transient regimes, which are highly nonlinear. A major and foundational open question is whether  there exists a device that realizes the time-reversal of a laser operating in {\it any} (generally highly-nonlinear) regime, i.e. capable of perfectly absorbs the field $E(t)$ emitted by a laser operating in any regime. If yes, how can we realize (at least in principle) such an "anti-laser" device?\\
 It is the aim of this work to answer to such two major questions. By considering an optical system with a single input/output channel and in the plane-wave approximation, we will prove rather generally that for any electromagnetic signal $E(t)$ generated by a laser system \textbf{S} operated in an arbitrary regime, i.e. emitting a  transient, irregular, chaotic, or periodic signal, one can always construct a CPA device \textbf{S'} that perfectly absorbs the field $E(t)$ emitted by \textbf{S}, and that  a possible simple realization of this system is obtained by placing inside \textbf{S}, nearby the output coupler, a broadband linear absorber (attenuator) of appropriate transmittance. As examples, we discuss two CPA devices that perfectly absorb the former a chaotic signal emitted by a single-mode laser operated in the Lorenz-Haken instability regime, the latter  a frequency-modulated (FM) optical wave emitted by a multimode FM-operated laser. 
 
 \section{Basic CPA idea}
 Let us consider a rather general ring-cavity laser system \textbf{S} of length $\mathcal{L}$ with a single output coupler, consisting of a lossless beam splitter BS, as shown in Fig.1(a).  The laser cavity contains a gain medium and possibly other optical elements or devices, such as saturable absorbers, amplitude or frequency modulators, etc. depending upon the operating regime of the laser. The electric field $\mathcal{E}(z,t)$ circulating inside the cavity can be written as $\mathcal{E}(z,t)=A(z,t) \exp(ikz-\omega t)$, where $\omega$ is a reference frequency, $k= \omega/c$ is the wave number in vacuum, $z$ is the longitudinal spatial coordinate along the ring, and $A(z,t)$ is a slowly-varying envelope. Without loss of generality, $\omega$ is chosen to be a resonance frequency of the empty cavity, so that $k \mathcal{L}$ is an integer multiple of $2 \pi$. After one cavity transit, the envelope $ \phi(t) \equiv A(z=\mathcal{L}^-,t)$ at the plane $z=\mathcal{L}^-$ can be formally written as 
 \begin{equation}
 \phi (t)=\hat{P} (\psi(t-T_R),t), 
 \end{equation}
where $\psi(t)=A(z=0^+,t)$ is the field envelope at the $z=0$ plane, $T_R=\mathcal{L}/c$ is the cavity transit time, and $\hat{P}$ is  a functional of $\psi(t-T_R)$  that describes the field evolution over one cavity round trip. The specific form of $\hat{P} (\psi,t)$ depends on the optical elements and devices put inside the cavity and in general can not be given explicitly, requiring to solve coupled equations that account for e.g. polarization and population dynamics in the active medium or in saturable absorbers possibly present inside the laser cavity with the appropriate initial conditions for the population and polarization variables (see, for instance, \cite{Newell}). Explicit forms of the operator $\hat{P}$ can be given is some cases, for example, for mode-locked lasers \cite{Haus}. However, for our purposes we do not need here to explicitly specify the form of $\hat{P}$.  The time delay $T_R$ in Eq.(1) is introduced in such a way that  for an empty cavity  $\hat{P}$ is the identity operator, i.e. $\hat{P}(\psi(t-T_R),t)=\psi(t-T_R)$. 
The evolution of $\psi(t)$ at successive cavity transits and the corresponding output laser field $E(t)$ are simply obtained by imposing the scattering relation between input/output channels at the beam splitter BS.  For a lossless BS, the scattering matrix is unitary, and one can write (see the inset of Fig.1)
\begin{eqnarray}
\left( \begin{array}{c}
\psi \\
E
\end{array}
\right)=
\left( \begin{array}{cc}
\sqrt{T} & \sqrt{R} \\
-\sqrt{R} & \sqrt{T}
\end{array}
\right)
\left( \begin{array}{c}
f \\
\phi
\end{array}
\right),
\end{eqnarray}
where $T$ and $R=1-T$ are the BS transmittance and reflectance, respectively, that are assumed to be spectrally flat. For the laser system \textbf{S} without an injected signal, one has $f(t)=0$,  $\phi(t)=(1/ \sqrt{R}) \psi(t)$ and thus one obtains the following equation for $\psi(t)$
\begin{equation}
\psi(t)=\sqrt{R} \hat{P}(\psi(t-T_R),t).
\end{equation}
The signal $E(t)$ emitted by the laser system in then given by
\begin{equation}
E(t) =\sqrt{\frac{T}{R}} \psi(t).
\end{equation}
Once the initial field distribution $\psi(t)$ is assigned in the interval $0<t<T_R$, Eq.(3) can be used to determine the evolution of $\psi(t)$ at successive transits in the ring. To clarify this point, let us introduce a local time variable $\tau$, with $0< \tau < T_R$, and let us set $t=nT_R+\tau$, where $n=0,1,2,3,...$ is the round trip number in the ring. After setting $\psi^{(n)}(\tau)=\psi(t=nT_R+\tau)$, Eq.(3) can be written as
\begin{equation}
\psi^{(n)}(\tau)= \sqrt{R} \hat{P} \left( \psi^{(n-1)}(\tau), \tau+nT_R \right).
\end{equation}
 For an assigned initial condition $\psi^{(0)}(\tau)$, the map (5) enables to determine recursively the field distributions $\psi^{(n)}(\tau)$ at successive round trips. The functional $\hat{P} \left( \psi^{(n-1)}(\tau), \tau+nT_R \right)$  is generally computed in the frame of an initial-boundary value (Goursat) problem of coupled differential equations  (see, for instance, \cite{Newell}; see also the example discussed in the Appendix).
\begin{figure}
\includegraphics[width=8.5cm]{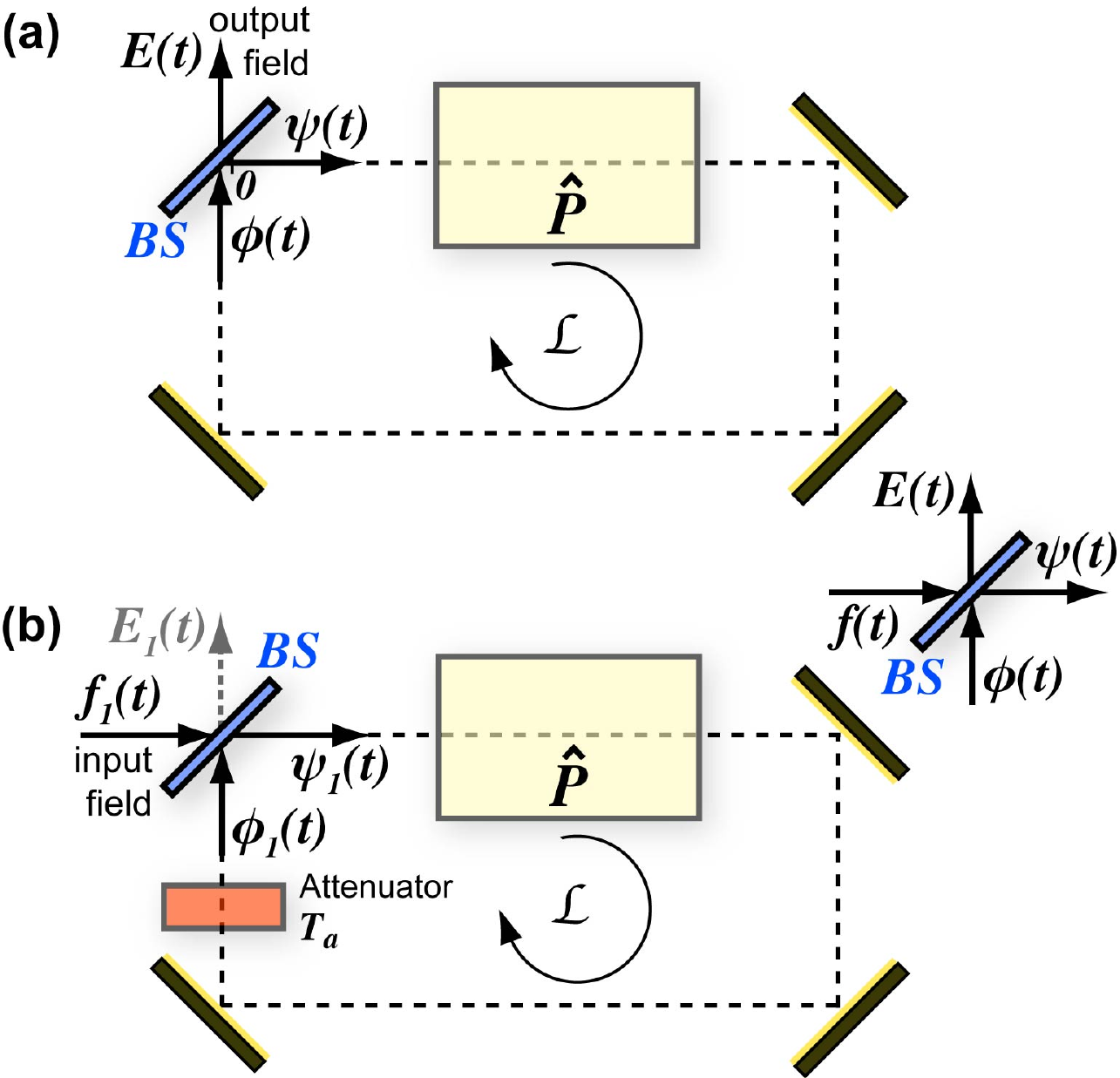}
\caption{(Color online) Schematic of (a) a laser system $\textbf {S}$, made of a ring-cavity of length $\mathcal {L}$ with a single output coupler (lossless beam splitter BS, with transmittance $T$ and reflectance $R=1-T$), and (b) corresponding CPA system \textbf {S'}. Note that \textbf{S'} is simply obtained from \textbf {S} by the insertion of a linear absorber (attenuator) nearby the output coupler with transmittance $T_a=R^2$. The inset in (a) shows the scattering relations of the field amplitudes at the lossless BS.}
\end{figure}
The temporal evolution of the emitted field $\psi(t)$ depends on the operating regime of the laser. For example, it could describe a periodic train of ultrashort pulses when the laser is operated in the mode-locking regime; in this case the self-consistent equation (3) has the form of a differential-delayed equation, the so-called master equation of mode-locking \cite{L3,Haus}. But it could also describe a rather irregular waveform, such as for a laser operated in the chaotic regime (see, for instance, \cite{L2,chaos}). The main question is now: is there an optical system \textbf{S'} that can perfectly absorb the optical field $E(t)$ emitted by the laser system \textbf{S}, i.e.~can we in some way time reverse a laser that is operated in an {\it arbitrary} regime? \\The answer is affirmative, and a  possible realization of the system \textbf{S'} is depicted in Fig.1(b). \textbf{S'} is basically obtained from \textbf{S} by just inserting, nearby the output beam splitter BS, a broadband linear absorber (attenuator) with a transmittance $T_a$ given by 
\begin{equation}
T_a=R^2. 
\end{equation}
 In this way, it can be shown rather generally that \textbf{S'} behaves as a CPA when the signal $f_1(t)$ injected into the cavity is given by
 \begin{equation}
 f_1(t)=\sqrt {R} E(t)=\sqrt{T} \psi(t)
 \end{equation}
 where $E(t)$ is given by Eq.(4) and $\psi(t)$ is the solution to Eq.(3) with the appropriate initial conditions. Such an injected signal is basically the output of the laser \textbf{S}, attenuated by the factor $\sqrt{R}$ \cite{note}.\par 
 To prove that \textbf{S'} time reverses the laser system \textbf{S} when Eqs.(6) and (7) are satisfied, let us inject into \textbf{S'} via the beam splitter port the signal 
 $f_1(t)$. Let us then indicate by $\mathcal{E}_1(z,t)=A_1(z,t) \exp(ikz-i \omega t)$ the electric field that is established inside the cavity of the system \textbf{S'} after signal injection, and by $E_1(t)$ the corresponding output field [see Fig.1(b)]. In this case one can obviously write [compare with Eq.(1)]
\begin{equation}
\phi_1(t)=\sqrt{T_a} \hat{P}( \psi_1(t-T_R),t)
\end{equation}
where we have set $\psi_1(t)=A_1(z=0^+,t)$ and $\phi_1(t)=A_1(z=\mathcal{L}^-,t)$. In fact, the cavity round-trip operator $\hat{P}_1$ of the system \textbf{S'} is simply obtained by cascading the propagator $\hat{P}$ of \textbf{S} with the transmission amplitude $\sqrt{T_a}$ of the added linear absorber, i.e. $\hat{P}_1= \sqrt{T_a} \hat{P}$. On the other hand, the scattering relation of the field amplitudes at the beam splitter BS in the system \textbf{S'} relates $\psi_1(t)$, $\phi_1(t)$ and the injected signal $f_1(t)$ according to
\begin{equation}
\psi_1(t)  =  \sqrt{T} f_1(t)+\sqrt{R} \phi_1(t).
\end{equation}
From Eq.(9) one obtains $\phi_1(t)=(1/  \sqrt{R}) \psi_1(t)- \sqrt{T/R} f_1(t)$, which after substitution into Eq.(8) yields the following equation for the intracavity field $\psi_1(t)$ that is established inside the cavity at the plane $z=0^+$
\begin{equation}
\psi_1(t)=\sqrt{T}f_1(t)+\sqrt{RT_a}  \hat{P}(\psi_1(t-T_R),t)
\end{equation}
The field $E_1(t)$ leaving $\textbf{S'}$ is then obtained from the scattering relation at the beam splitter and reads explicitly
$E_1(t)  =  -\sqrt{R}f_1(t)+\sqrt{T} \phi_1(t)$, which by means of Eq.(9) can be written as
\begin{equation}
E_1(t)=\sqrt{\frac{T}{R}} \psi_1(t)-\frac{f_1(t)}{\sqrt{R}}.
\end{equation}
Similarly to the case of the laser system $\mathbf{S}$ in the absence of the injected signal, Eq.(10) can be written as a recursive relation
\begin{equation}
\psi_1^{(n)}(\tau)=\sqrt{T}f_1^{(n)}(\tau)+\sqrt{RT_a} \hat{P}(\psi_1^{(n-1)}(\tau), \tau+nT_R)
\end{equation}
where we have set $\psi_1^{(n)}(\tau)=\psi_1(\tau+nT_R)$ and $f_1^{(n)}(\tau)=f_1(nT_R+\tau)$. Once the initial field distribution $\psi_1^{(0)}(\tau)$ is assigned, the map (12) enables one to determine the evolution of the intracavity field $\psi_1(t)$ at successive round-trips. Let us now assume that the two systems $\mathbf{S}$ and $\mathbf{S}'$ are initially prepared in the same state, i.e. that $\psi_1^{(0)}(\tau)=\psi^{(0)}(\tau)$ and that atomic variables (population inversion and polarization) have the same initial values in the active medium (for example their equilibrium values).  In this case, one can readily show using Eq.(5) that, if the injected signal $f_1(t)$ is chosen according to
\begin{equation}
f_1(t)=\frac{1-\sqrt{T_a}}{\sqrt{T}}\psi(t)
\end{equation}
the recursive relation (12) admits of the solution $\psi_1^{(n)}(\tau)=\psi^{(n)}(\tau)$ for any $n$.  This implies that the intracavity fields $\psi_1(t)$ and $\psi(t)$ established in the two systems $\mathbf{S}$ and $\mathbf{S}'$ are the same. 
 Note that this result holds even if the map (12)  admits of different attractors (such as for a bistable system), provided that the initial conditions in  $\mathbf{S}'$ belong to the basin of attraction of the CPA solution $\psi_1^{(n)}(\tau)=\psi^{(n)}(\tau)$. This ensures that asymptotically one has $\psi_1^{(n)}(\tau) \rightarrow \psi^{(n)}(\tau)$ as $n \rightarrow \infty$, even though $\psi_1^{(n)}(0) \neq \psi^{(n)}(0)$. But for some other initial conditions it might happen that $\psi_1^{(n)}(\tau)$ is not attracted toward $\psi^{(n)}(\tau)$, and the CPA scheme fails. However, like in any nonlinear dynamical system showing different attractors, one can switch the path from one stable attractor to another one by introducing large perturbations of the system parameters (e.g. by transiently changing the cavity losses or the gain in the medium). As discussed in Sec.III.A with reference to a specific example , the result holds for transient or chaotic regimes as well, which are very sensitive to the initial state of the system.\\  The field $E_1(t)$ emitted by the system $\mathbf{S}'$ is obtained after substitution of Eq.(13) into Eq.(11). For $\psi_1(t)=\psi(t)$,  one obtains
 \begin{equation}
E_1(t)=\frac{\sqrt{T_a}-R}{\sqrt{TR}} \psi(t)
\end{equation}
which vanishes if the transmittance $T_a$ is chosen to satisfy Eq.(6). Correspondingly, the injected field $f_1(t)$, as obtained from Eq.(13) after setting $T_a=R^2$, is given by Eq.(7). Hence, provided that Eqs.(6) and (7) are met and the two systems $\mathbf{S}$ and $\mathbf{S}'$ are initially prepared in the same state, $\mathbf{S}'$ behaves as a CPA device for the signal emitted by $\mathbf{S}$, regardless of its operational regime.
  
\section{Examples of CPA devices}
In this section we discuss two examples of CPA devices corresponding to a laser oscillator $\mathbf{S}$ operating in two nontrivial regimes. The former example is a CPA device for a chaotic optical signal emitted by a single-mode homogeneously-broadened laser  operated in the chaotic (Lorenz-Haken instability) regime \cite{L1,L2,chaos}; the latter example is  a CPA device for a frequency-modulated (FM) optical signal emitted by   a FM-operated laser with an intracavity phase modulator \cite{L3,FM1,FM2}.
\subsection{CPA for a chaotic optical field}
\begin{figure}
\includegraphics[width=8.2cm]{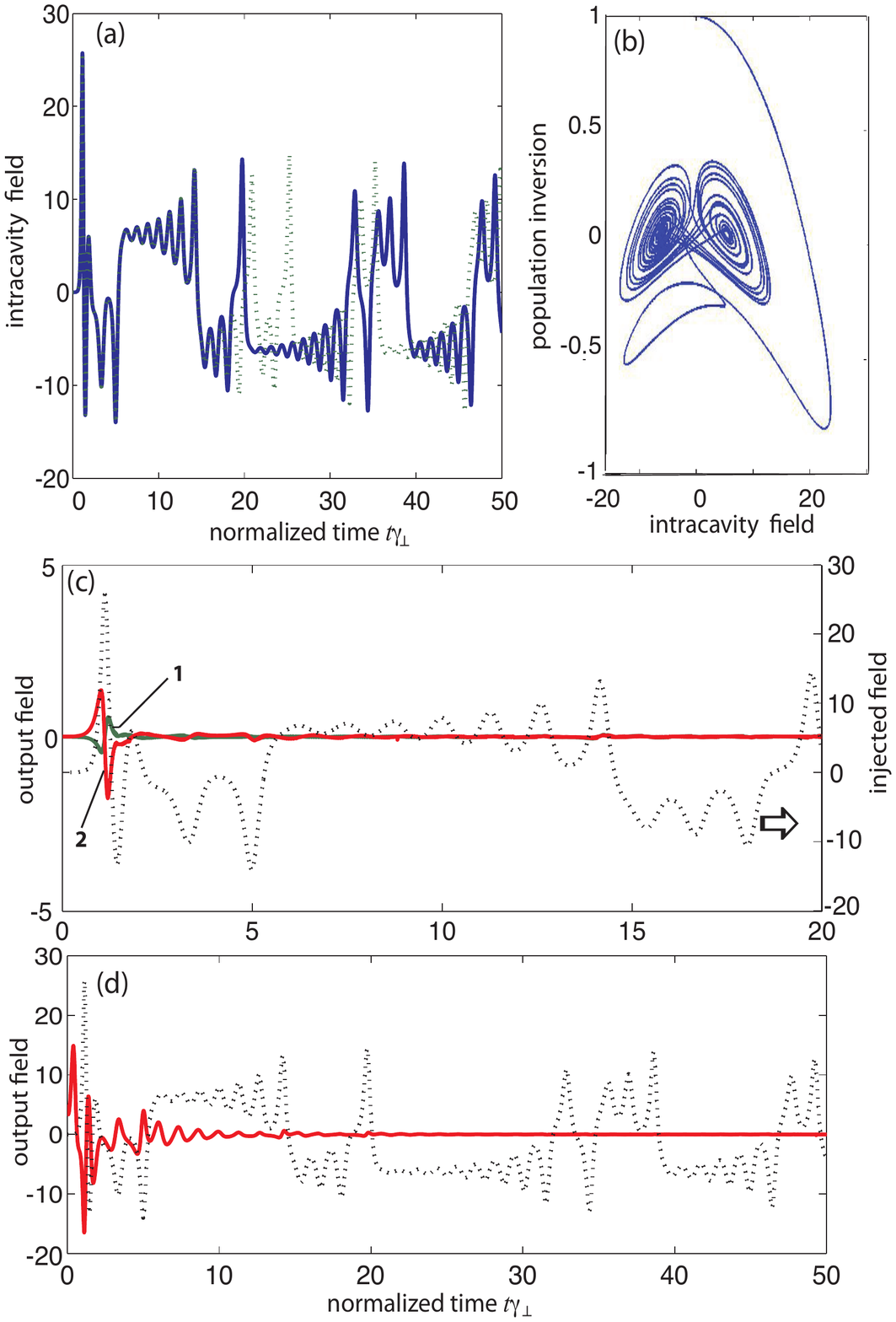}
\caption{(Color online) (a) Evolution of the intracavity field $\psi(t)$ in the laser system $\mathbf{S}$ as obtained by numerical solution of the Lorenz-Haken equations (15) for the initial condition $(\psi(0)=0.001,\Lambda(0)=0,n(0)=1)$ (solid curve), and (b) corresponding projection of the phase-space trajectory in the $(\psi,n)$ plane. Parameter values are $\kappa/ \gamma_{\perp}=4$, $\gamma_{\parallel} / \gamma_{\perp}=0.5$, $2C=40$ and $\Delta=0$. In (a) the dotted curve shows the evolution of $\psi(t)$ as obtained by the slightly different initial condition $(\psi(0)=0.0011,\Lambda(0)=0,n(0)=1)$. Note the strong sensitivity on the initial condition, which is a rather general feature of a chaotic attractor. In (c) we show the numerically computed evolution of the output field $E_1(t)$,  normalized to $\sqrt{T}$ (i.e. the behavior of $\psi_1(t)-\psi(t)$), in the CPA system $\mathbf{S}'$ with an injected field $f_1(t)$ and attenuator transmittance $T_a$ satisfying Eqs.(6) and (7), for the initial conditions $(\psi_1(0)=0.0001,\Lambda_1(0)=0,n_1(0)=1)$ (curve 1), and $(\psi_1(0)=0.005,\Lambda_1(0)=0,n_1(0)=1)$ (curve 2). The dotted curve in (c) shows, for comparison, the behavior of the injected field $f_1(t)$, normalized to $\sqrt{T}$ [i.e. the behavior of $\psi(t)$]. (d) Same as (c), but for a large deviation of initial conditions $(\psi_1(0)=5,\Lambda_1(0)=0,n_1(0)=1)$.}
\end{figure}
The first example we would like to discuss is a CPA device that perfectly absorbs a chaotic optical field emitted by a  single-mode homogeneously-broadened laser  operated in the chaotic (Lorenz-Haken instability) regime \cite{L1,L2,chaos}. In this case, the laser system $\mathbf{S}$ just contains a homogeneously-broadened two-level medium of length $l$ with a small-signal gain coefficient per unit length $g$, which is provided by population inversion in the medium. In the single longitudinal mode and uniform field approximations, the intracavity field $A(z,t)$ is assumed to be almost uniform along the ring (i.e. almost constant for $0<z< \mathcal{L}$) and slowly-varying in time over one cavity round-trip time \cite{L2}.  In this case, the map (5) can be effectively replaced by a set of three coupled differential equations describing the evolution of the intracavity field, polarization and population inversion in the two-level medium, the so-called Lorenz-Haken model of the single-mode homogeneously-broadened laser (see the Appendix for technical details). These equations read \cite{L2}
\begin{eqnarray}
\frac{d \psi}{dt} & = & -\kappa( \psi+2C \Lambda) \nonumber \\
\frac{d \Lambda}{dt} & = & -\gamma_{\perp}[(1+i \Delta) \Lambda + \psi n] \\
\frac{dn}{dt}& = & -\gamma_{\parallel} \left[ n-1 -\frac{1}{2} ( \psi \Lambda^*+\psi^* \Lambda) \right] \nonumber
\end{eqnarray}
where: $\psi(t)$, $\Lambda(t)$ and $n(t)$ are the intracavity electric field, polarization and population inversion, respectively, normalized as in Refs.\cite{L2,uffa3};  $\gamma_{\parallel}$ and $\gamma_{\perp}$ are the
population and dipole decay rates, respectively;
\begin{equation}
\kappa=\frac{Tc}{2 \mathcal{L}}
\end{equation}
 is the cavity decay rate; $C=gl/T$; and $\Delta=(\omega_0-\omega) / \gamma_{\perp}$ is the normalized detuning parameter between the cavity resonance frequency $\omega$ and the atomic transition frequency $\omega_0$. 
In particular, at resonance $\Delta=0$ (i.e. for $\omega=\omega_0$), $\Lambda$ and $\psi$ can be taken to be real-valued, and Eqs.(15) are analogous to the Lorenz model, developed for convective instabilities in hydrodynamics \cite{L1,L2}. In the following, we will mainly limit to consider the limit $\Delta=0$. 
In this case, laser threshold is attained for $2C>1$, at which the non-lasing trivial solution ($\psi=0,\Lambda=0,n=1$) to Eqs.(15) becomes unstable.  Above threshold, the laser equations admit of a steady-state solution, with two possible phases for $\psi$ and $\Lambda$ ($0$ and $\pi$), namely
\begin{equation}
\psi_{\pm}= \pm \sqrt{2C-1} , \; \Lambda_{\pm}= \mp \frac{1}{2C} \sqrt{2C-1}, \; n_{\pm}=\frac{1}{2C}.
\end{equation}
Such solutions may undergo a subcritical Hopf instability in the bad cavity limit ($\kappa> \gamma_{\parallel}+\gamma_{\perp}$) and for large small-signal gains \cite{L1,L2}. In such a regime chaotic self-pulsations can be observed. Here, the point ($\psi(t)$, $\Lambda(t)$, $n(t)$) in phase space never settles down but continually makes excursions about one of the two laser solutions (17) with what appear to be random jumps from circling one fixed point to circling the other. The strange set is known as the Lorenz attractor, and the output laser field $E(t) \simeq \sqrt{T} \psi(t)$ turns out to be strongly sensitive to the initial conditions \cite{L1,L2}. As an example, in Fig.2(a) we show the evolution of the intracavity laser field $\psi(t)$ as obtained by numerical solution of the Lorenz-Haken equations (15) for parameter values $\kappa / \gamma_{\perp}=4$, $\gamma_{\parallel}/ \gamma_{\perp}=0.5$, $2C=40$, $\Delta=0$ and for the initial condition $\psi(0)=0.001$, $\Lambda(0)=0$, $n(0)=1$. The corresponding projection of the phase-space trajectory in the $(\psi,n)$ plane is depicted in Fig.2(b), which shows the characteristic Lorenz attractor of the laser dynamics in the chaotic regime. It should be noted that in such a regime the established intracavity laser field (and hence the laser output field) is strongly sensitive to the initial conditions. For example, the dotted curve in Fig.2(a) shows the intracavity field that one would observe for the slightly changed initial conditions  $\psi(0)=0.0011$, $\Lambda(0)=0$, $n(0)=1$.\par
Let us now consider the CPA device $\mathbf{S}'$ associated to $\mathbf{S}$, as discussed in Sec.II. In this case, in the single-longitudinal mode and mean-field approximations, the dynamical evolution of the intracavity field $\psi_1(t)$ is governed by a set of coupled equations similar to Eqs.(15), in which the first equation is modified to take into account the effects of the injected field $f_1(t)$ and of the intracavity absorber with transmittance $T_a$. One obtains (see the Appendix for technical details)
\begin{eqnarray}
\frac{d \psi_1}{dt} & = & -\kappa( \psi_1+2C \Lambda_1) -\kappa_1 \psi_1 + \kappa \frac{2f_1(t)}{\sqrt{T}} \nonumber \\
\frac{d \Lambda_1}{dt} & = & -\gamma_{\perp}[(1+i \Delta) \Lambda_1 + \psi_1 n_1) \\
\frac{dn}{dt}& = & -\gamma_{\parallel} \left[ n-1 -\frac{1}{2} ( \psi \Lambda^*+\psi^* \Lambda) \right] \nonumber
\end{eqnarray}
where
\begin{equation}
\kappa_1=\frac{c(1-\sqrt{T_a})}{\mathcal{L}}=2 \frac{(1-\sqrt{T_a})}{T} \kappa.
\end{equation}
The output field $E_1(t)$ from the CPA device is then given by Eq.(11). 
According to the general analysis presented in Sec.II, if the injected field is chosen to satisfy the condition (13), it can be readily shown that the solution to Eqs.(18) is given by $\psi_1(t)=\psi(t) $, $\Lambda_1(t)=\Lambda(t) $, $n_1(t)=n(t)$ {\it provided that} the same initial conditions $\psi_1(0)=\psi(0)$, $\Lambda_1(0)=\Lambda(0)$, $n_1(0)=n(0)$ are assumed for the field and atomic variables in the two systems $\mathbf{S}$ and $\mathbf{S}'$. Furthermore, if the transmittance $T_a$ of the absorber in $\mathbf{S}'$ is tuned to satisfy the condition (6), the output field $E_1(t)$ emitted by $\mathbf{S}'$ vanishes, i.e. $\mathbf{S}'$ behaves as a perfect CPA for the chaotic optical field $\psi(t)$. Note that for the perfect CPA device one has $\kappa_1=2 \kappa$ and the dynamical system (18) reads 
\begin{eqnarray}
\frac{d \psi_1}{dt} & = & -\kappa( \psi_1+2C \Lambda_1) - 2 \kappa [\psi_1 -\psi(t)] \nonumber \\
\frac{d \Lambda_1}{dt} & = & -\gamma_{\perp}[(1+i \Delta) \Lambda_1 + \psi_1 n_1) \\
\frac{dn_1}{dt}& = & -\gamma_{\parallel} \left[ n_1-1 -\frac{1}{2} ( \psi_1 \Lambda_1^*+\psi_1^* \Lambda_1) \right] \nonumber
\end{eqnarray}
where $\psi(t)$ is the solution to Eqs.(15) with the assigned initial conditions.\\ 
 It should be noted that in a practical case the laser system $\mathbf{S}$ and the CPA system $\mathbf{S}'$ can not be exactly prepared in the same state. In this case,
 the solution $(\psi_1, \Lambda_1,n_1)$ to Eqs.(20)  deviates from the solution $(\psi,\Lambda,n)$ of Eqs.(15), just because the initial conditions of field and/or atomic variables are different. Correspondingly, the output field $E_1(t)$ emitted by the CPA  device, given by
\begin{equation}
E_1(t)=\sqrt{\frac{T}{R}}[\psi_1(t)-\psi(t)] \simeq \sqrt{T}[\psi_1(t)-\psi(t)] ,
\end{equation}
would not vanish, i.e. perfect absorption would be lost. Such a 
circumstance rises the question whether the CPA idea presented in Sec.II is actually of physical relevance. This objection is especially a serious one whenever the system dynamics is strongly sensitive to the initial conditions, such as for a chaotic laser, or if the system $\mathbf{S}'$ with injected signal may show two (or more) stable attractors, i.e. in the presence of bistability or multistability \cite{note2}. In this case, an initial condition of $\mathbf{S}'$ sufficiently far from that of $\mathbf{S}$ can bring the dynamical system (20)  into a {\it different attractor} than $(\psi(t), \Lambda(t), n(t))$ \cite{note2}. Strong deviations of the initial conditions may occur, for example, whenever the pump parameter $2C$ is larger than $3$ and system $\mathbf{S}'$ thus self-oscillates, i.e. it is above laser threshold (in the absence of the injected signal) in spite of the attenuator put in the cavity.  However, provided that the basin of attraction of $(\psi(t), \Lambda(t), n(t))$ for $\mathbf{S}'$ is sufficiently wide, a {\it large perturbation} can  switch the path of the dynamical system toward the "right" CPA attractor $(\psi(t), \Lambda(t), n(t))$.  Of course the basin of attraction of  $(\psi(t), \Lambda(t), n(t))$ for $\mathbf{S}'$, as well as the kind and strength of the large perturbations requested to switch the nonlinear dynamics of Eqs.(20) from one attractor to another one, should be considered on a case-by-case basis. Let us focus here our attention to the Lorenz-Haken model of laser chaos. In this regime it is known that even a small change in the value of $\psi(0)$ may deeply modify the output waveform, as shown in Fig.2(a). However, numerical results show that the basin of attraction of  $(\psi(t),\Lambda(t),n(t))$ for Eqs.(20) is quite broad, i.e. asymptotically one has $\psi_1(t) \rightarrow \psi(t)$ for a quite broad  range of initial conditions $(\psi_1(0),\Lambda_1(0),n_1(0))$ around  $(\psi(0),\Lambda(0),n(0))$. The reason thereof is that the dynamical system (20) [contrary to the Lorenz-Haken dynamical system (15)] is {\it driven} by the external field $\psi(t)$, which forces $\psi_1(t)$ toward $\psi(t)$ after an initial transient. This is clearly shown in Fig.2(c), which depicts the output field $E_1(t)$, normalized to $\sqrt{T}$ [i.e. the difference $\psi_1(t)-\psi(t)$; see Eq.(21)]  emitted by the CPA device $\mathbf{S}'$, as obtained by numerical integration of Eqs.(20), for the two initial  conditions ($\psi_1(0)=0.0001,\Lambda_1(0)=0$, $n_1(0)=1$) [curve 1 in Fig.2(c)] and ($\psi_1(0)=0.005,\Lambda_1(0)=0$, $n_1(0)=1$) [curve 2 in Fig.2(c)], which appreciably differ from $\psi(0)=0.001,\Lambda(0)=0$, $n(0)=1$. For comparison, in the figure we show (but on a different scale) the field $f_1(t)=\sqrt{T} \psi(t)$ injected into the cavity, normalized to $\sqrt{T}$ (i.e. the behavior of $\psi(t)$), where $\psi(t)$ is the field of $\mathbf{S}$ for the initial condition ($\psi(0)=0.001,\Lambda(0)=0$, $n(0)=1$) [i.e. the solid curve in Fig.2(a)]. Note that, since the initial conditions in $\mathbf{S}'$ and $\mathbf{S}$ are different, the CPA output field does not exactly vanish in an initial transient [see curves 1 and 2 in Fig.2(c)], however it remains much smaller than the injected field even though the initial conditions in the two systems are appreciably different. Most importantly, after an initial transient the output field vanishes, and perfect absorption is attained. Such a behavior is observed even for large deviations of initial conditions, as shown in Fig.2(d) as an example.  

\subsection{CPA for a frequency-modulated optical field}
As a second example, we discuss the realization of a CPA device for an optical frequency-modulated signal $E(t)$, i.e. an optical field with a constant intensity but with a sinusoidally-modulated optical phase. A coherent FM optical signal $E(t)$ is approximately generated by placing a phase modulator inside a laser device and sinusoidally-driving it asynchronously as compared to the cavity round-trip time. In this way the laser operates in the so-called frequency-modulation (FM) regime \cite{L3,FM1,FM2}. The FM regime is a multimode regime in which several cavity axial modes are excited out of resonance with steady-state amplitudes but with time-varying phases, resulting in a constant intensity laser field but with a carrier laser frequency which is sinusoidally swept at the frequency impressed by the phase modulator \cite{L3}.  
As shown in Ref.\cite{FM2},  a pure FM signal at optical carrier can be generated provided that the phase modulation is sufficiently asynchronous such that the bandwidth of the FM signal is much smaller that the  gain bandwidth of the active medium.  
A schematic of a FM-operated laser is shown in Fig.3(a). The optical cavity contains a gain medium and a phase modulator, which is sinusoidally-driven at a frequency $\omega_m$ which is enough detuned from the cavity axial mode separation $ \omega_{ax}=2 \pi / T_R$, where $T_R$ is the cavity photon transit time \cite{L3}.  For a slow gain medium and neglecting dispersive and finite gain bandwidth effects, the cavity round-trip operator reads \cite{FM2}
\begin{equation}
\hat{P}(t,\psi)=\exp(g) \exp[i \Delta \cos(\omega_m t)] \psi(t-T_R),
\end{equation}
where $g$ is the single-pass saturated gain in the active medium and $\Delta$ is the modulation depth impressed by the phase modulator.  The self-consistent equation for the field $\psi(t)$ [Eq.(3)] then  reads
\begin{equation}
\psi(t)=\exp[g-\gamma+i\Delta \cos(\omega_m t)] \psi(t-T_R),
\end{equation}
where  $\gamma = -{\rm ln} \sqrt{R}$ is the logarithmic loss of the cavity due to the output coupling.  Equation (23) should be supplemented with a rate equation for the saturated gain $g$ (see, for instance, \cite{ufff}). After an initial transient, a steady-state operation is achieved, in which the saturated gain settles down to the stationary value  $g=\gamma$ \cite{ufff}.Correspondingly, the intracavity field $\psi(t)$ is given by one of the "modes" of the phase-modulated cavity (generally the one with the carrier fequency closest to the center of the gainline). Such modes are simply obtained as eigenfunctions of Eq.(23) and reads explicitly \cite{FM2,ufff}
\begin{equation}
\psi(t)=\exp[i \Gamma \cos(\omega_m t +\varphi)+il \omega_{ax}t].
\end{equation}
In Eq.(24), $l=0, \pm1, \pm 2, ...$ is the mode index,  $\Gamma$ is the effective modulation index, and $\varphi$ a phase offset. The values of $\Gamma$ and $\varphi$ are obtained after substitution of the Ansatz (24) into Eq.(23) and equating the imaginary terms in the exponentials on the left and right hand sides of the equation so obtained. This yields
\begin{eqnarray}
\Gamma & = & \frac{\Delta}{2  \sin ( \pi \omega_m / \omega_{ax})} \\
 \cos \varphi & = & \frac{\Delta}{2 \Gamma}.
 \end{eqnarray}
  The output field,  $ \sim \psi(t)e^{i\omega t}$, is thus a pure FM signal, i.e.~a frequency comb, with spectral lines at frequencies $n\omega_m$ ($n=0, \pm 1, \pm 2,...$) around the carrier frequency $\omega+l \omega_{ax}$ of amplitudes $\sim J_n(\Gamma)$, where $J_n$ is the Bessel function of first kind and of order $n$, and $\omega$ is a reference cavity resonance frequency.  
According to Eq.(25), the effective modulation index $\Gamma$ increases and diverges as the synchronous modulation condition $\omega_m= \omega_{ax}$ is attained. Indeed, in practical FM-operated lasers large modulation indices are usually obtained by tuning the modulation frequency $\omega_m$  close to the cavity axial mode spacing $\omega_{ax}$ (see, for instance, \cite{FM1,FM3}). However, the solution (24) provides an accurate approximation to the output field of the FM-operated  laser provided that the modulation frequency $\omega_m$ remains sufficiently detuned far apart from $\omega_{ax}$ in such a way that the spectral extent $\sim 2 \Gamma \omega_m$ of the FM signal $\psi(t)$ is much smaller than the gain linewidth $\omega_g$, i.e. provided that the following condition 
\begin{equation}
|\omega_m-\omega_{ax}| \gg \frac{\omega_{ax} \omega_m  \Delta}{\pi \omega_g}
\end{equation}
is satisfied. 
\begin{figure}
\includegraphics[width=7cm]{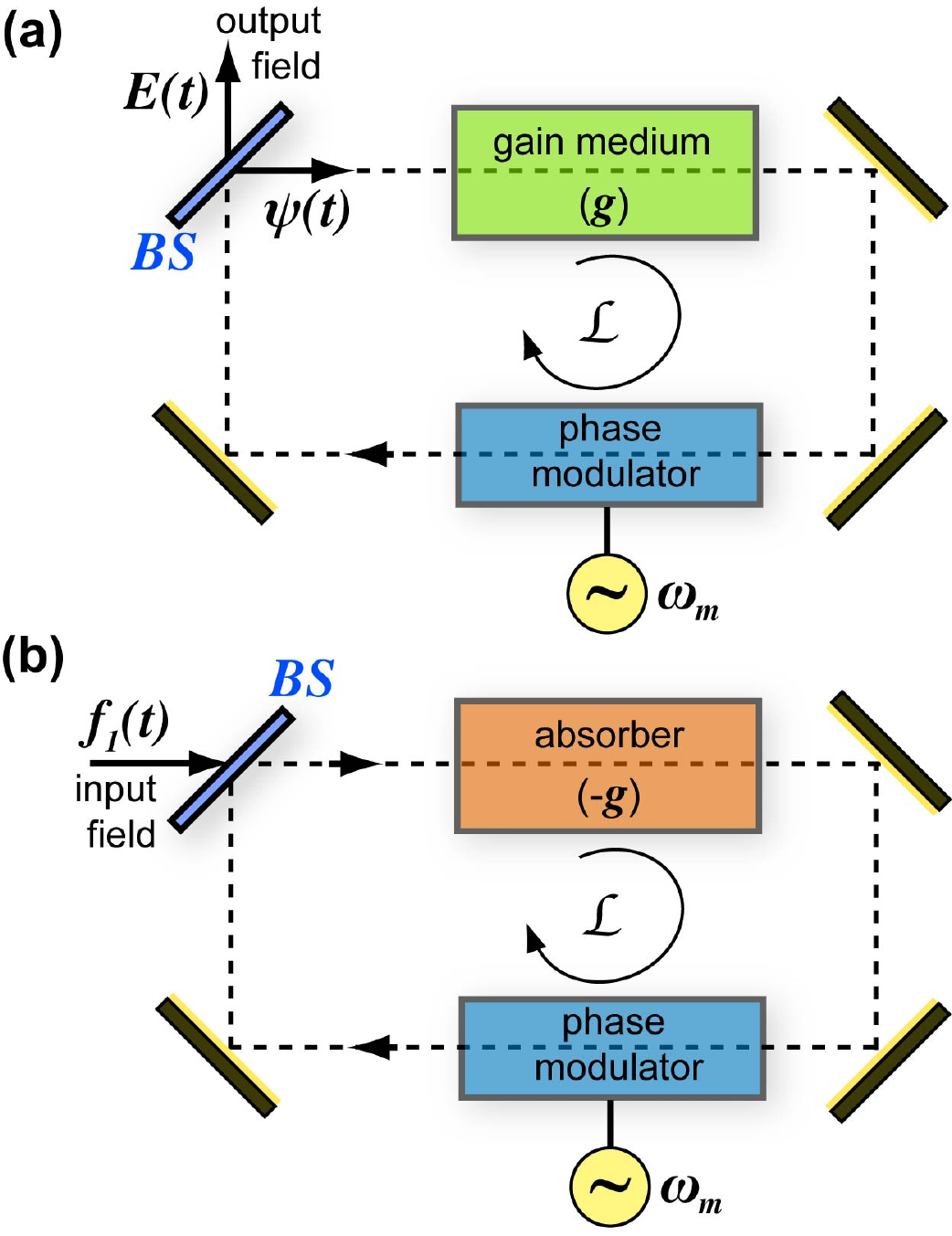}
\caption{(Color online) Schematic of (a) a FM-operated laser \textbf {S}, made of a ring-cavity containing a gain medium (gain coefficient $g$) and a phase modulator, driven at a frequency $\omega_m$ detuned from the frequency separation $\omega_{ax}$ of cavity axial modes, and (b) the corresponding CPA system \textbf {S'}, which is obtained from \textbf{S} by replacing the gain medium with an absorber with absorption coefficient $-g$.}
\end{figure}
 Indeed, as the synchronous modulation condition $\omega_m= \omega_{ax}$ is approached, laser operation switches into a pulsed regime (FM mode-locking), with the generation of a periodic train of short optical pulses  \cite{L3,FM4}.\par
Let us now discuss the possibility to perfectly absorb the FM signal emitted by $\mathbf{S}$. According to the analysis of Sec.II, a CPA system \textbf{S'} that perfectly absorbs  the FM signal can be obtained from \textbf{S} by placing inside the cavity a linear absorber with a transmittance $T_a=R^2$.  Note that, as $g=\gamma= - {\rm ln} \sqrt{R}$ and since we neglected gain bandwidth and dispersion effects, the combined effect of the gain and absorber media in \textbf{S'} is equivalent to that of a single linear absorber placed in the cavity with an absorption coefficient $-g$, as shown in Fig.3(b). In other words, for this special example  we retrieve the simple rule of Ref.\cite{Stone1}  that the CPA system is obtained from the lasing system by replacing the gain medium with an absorber with an amount of dissipation exactly opposite to the amplification factor in the lasing medium at threshold. We stress, however, that the CPA system \textbf{S'} that realizes the time reverse of the lasing system \textbf{S}, obtained by the general procedure outlined in Sec.II, does not generally correspond to the replacement of the gain medium with an absorber, as discussed for instance in the example of Sec.III.A. 

\section{Conclusion}
In conclusion, in this work we have extended the concepts of time-reversal of lasing and CPA, recently introduced in Ref.\cite{Stone1},  to the case of optical radiation emitted by  a laser operated in an arbitrary regime, i.e. for transient, chaotic or periodic  coherent optical fields.  We have proven rather generally that any electromagnetic signal $E(t)$ generated by a laser system \textbf{S} operated in an arbitrary (and generally highly-nonlinear) regime  can be perfectly absorbed by a CPA device $\bf{S'}$  which is simply realized by placing inside \textbf{S} a broadband linear absorber (attenuator) of appropriate transmittance. As examples, we discussed CPA devices that perfectly absorb a chaotic laser signal and a frequency-modulated optical wave.

\acknowledgments

This work was supported by the Italian MIUR (Grant No. PRIN-20082YCAAK, "Analogie ottico-quantistiche in strutture
fotoniche a guida d'onda") and by the Fondazione Cariplo (Grant No. 2011-0338, "New Frontiers in Plasmonic Nanosensing").

\appendix
\section{CPA for chaotic optical fields: derivation of the mean-field equations}
In this section we briefly derive the mean-field equations (15) and (18) given in the text from the general maps (5) and (12) in case where the functional $\hat{P}$ describes field propagation in a homogeneously-broadened two-level medium of length $l$. We will perform such a derivation for the more general case of the system $\mathbf{S}'$ of Fig.1(b) with arbitrary injected signal $f_1(t)$ and transmittance $T_a$ of the absorber;  the laser system $\mathbf{S}$ of Fig.1(a) is simply obtained after setting $f_1=0$, $T_a=0$. 
Field propagation inside the ring cavity is described by the Maxwell-Bloch equations (see Sec.4.3 of Ref.\cite{L2}; see also Ref.\cite{uffa3})
\begin{eqnarray}
\frac{\partial A_1}{\partial z} & = & - \frac{1}{c} \frac{\partial A_1}{\partial t}  -g \Lambda_1 \nonumber \\
\frac{\partial \Lambda_1}{\partial t} & = & - \gamma_{\perp} \left[ (1+i\Delta) \Lambda_1+n_1 A_1 \right] \\
\frac{\partial n_1}{\partial t} & = & - \gamma_{\parallel} \left[ n_1-1-\frac{1}{2} \left( A_1^* \Lambda_1+A_1 \Lambda_1^* \right) \right] \nonumber
\end{eqnarray}
where: $A_1(z,t)$, $\Lambda_1(z,t)$, $n_1(z,t)$ are the intracavity electric field envelope, polarization envelope and population inversion, respectively, normalized as in Refs.\cite{L2,uffa3}; $\gamma_{\parallel}$ and $\gamma_{\perp}$ are the
population and dipole decay rates, respectively;  $g$ is the small-signal gain per unit length in the active medium; and $\Delta=(\omega_0-\omega) / \gamma_{\perp}$ is the normalized detuning parameter between the cavity resonance frequency $\omega$ and the atomic transition frequency $\omega_0$. For the sake of definiteness, the gain medium (of lenght $l$) is placed at $0<z<l$; similarly, we assume that the attenuator $T_a$ is a thin plate placed close to the beam splitter BS [see Fig.1(b)]. In this way field attenuation after passage across the absorber can be included in the scattering matrix of the beam splitter BS. This yields the following ring-cavity boundary condition for the field $A_1(z,t)$ at 
the planes $z=0^+$ and $z=\mathcal{L}^-$
\begin{equation}
A_1(0^+,t)= \sqrt{T}f_1(t)+ \sqrt{RT_a}A_1(\mathcal{L}^-,t).
\end{equation}
The output field $E_1(t)$ escaping from the cavity through the beam splitter BS is then given by
\begin{equation}
E_1(t)=\sqrt{\frac{T}{R}} A_1(0^+,t)-\frac{f_1(t)}{\sqrt{R}}.
\end{equation}
Obviously $n_1$ and $\Lambda_1$ in Eqs.(A1) are defined in the range $0<z<l$, and for $z>l$ (i.e. in the empty cavity region) the first equation of (A1) is still valid provided that we assume $g=0$.
The field $A_1(\mathcal{L}^-,t)$ at $z=\mathcal{L}^-$ and time $t$ entering in  Eq.(A2) can be computed as a functional of the field $A_1(0^+,t-T_R)$ at plane $z=0^+$ and time $t-T_R$, where $T_R=\mathcal{L}/c$ is the cavity transit time. In fact, after introduction of the new variables
\begin{equation}
\xi=z \; , \; \eta=t-z/c
\end{equation}
Eqs.(A1) take the form
\begin{eqnarray}
\frac{\partial A_1}{\partial \xi} & = &  -g \Lambda_1 \nonumber \\
\frac{\partial \Lambda_1}{\partial \eta} & = & - \gamma_{\perp} \left[ (1+i\Delta) \Lambda_1+n_1 A_1 \right] \\
\frac{\partial n_1}{\partial \eta} & = & - \gamma_{\parallel} \left[ n_1-1-\frac{1}{2} \left( A_1^* \Lambda_1+A_1 \Lambda_1^* \right) \right] \nonumber
\end{eqnarray}
Equations (A5) can be integrated in the interval $0< \xi <l$ and for $0<\eta<T_R$ once the initial conditions $A_1(\xi=0^+, \eta)$ and $\Lambda_1(\xi,\eta=0)$,  $n_1(\xi,\eta=0)$ are assigned (initial-boundary value (Goursat) problem \cite{Newell}). Typically, for a medium initially at equilibrium in the absence of the field, one can take $ \Lambda_1(\xi,\eta=0)=0$ and $n_1(\xi, \eta=0)=1$. In this way, one can compute $A(\xi=\mathcal{L}^-, \eta)=A(\xi=l, \eta)$ for $0<\eta< T_R$. In the physical space-time variables $(z,t)$, this means that we know the value of $A_1(\mathcal{L}^-,t)$ in the time interval $T_R<t<2T_R$.  With such a solution, from Eq.(A2) one can then calculate $A_1(0^+,t)$ is the time interval $T_R<t<2 T_R$, which serves as an initial condition to integrate Eqs.(A5) in the interval $0<\xi<l$ and for $ T_R<\eta< 2 T_R$ (second round-trip) with the appropriate initial values of $\Lambda_1(\xi, \eta=T_R)$ and $n_1(\xi, \eta=T_R)$ computed at the previous step. Iteration of such a procedure enables one to calculate $A_1(0^+,t)= \psi_1(t)$ at any time $t>0$, and thus the output field $E_1(t)$ according to Eq.(A3). \par
An important case is the one obtained by taking the limit $gl \rightarrow 0$. In this case, from Eqs.(A5) it follows that the variables $A_1$, $\Lambda_1$ and $n_1$ are almost constant functions with respect to the $\xi$ variable, i.e. $A_1(\xi, \eta) \sim A_1(0,\eta)$, $\Lambda_1(\xi, \eta) \sim \Lambda_1(0,\eta)$ and $n_1(\xi, \eta) \sim n_1(0,\eta)$. The small change $A_1(\xi=\mathcal{L}^-, \eta) - A_1(\xi=0^+,\eta)$ can be computed  from the first of Eqs.(A5) as 
$A_1(\xi=\mathcal{L}^-, \eta) - A_1(\xi=0^+,\eta) \simeq - gl \Lambda_1(\xi=0,\eta)$, which in terms of the space-time physical variables $(z,t)$ reads
\begin{equation}
A_1(\mathcal{L}^-,t+T_R) \simeq A_1(0^+,t) - gl \Lambda_1(0,t).
 \end{equation}
Indicating by $\psi_1(t)=A_1(z=0^+,t)$, $\Lambda_1(t)=\Lambda_1(z=0,t)$ and  $n_1(t)=n_1(z=0,t)$, from Eqs. (A2) and (A6) it follows that $\psi_1(t)$ satisfies the delayed equation
\begin{equation}
\psi_1(t+T_R)=\sqrt{T} f_1(t+T_R)+ \sqrt{RT_a}[\psi_1(t)-gl \Lambda_1(t)]
\end{equation}
which should be associated to the differential equations for $\Lambda_1(t)$ and $n_1(t)$
\begin{eqnarray}
\frac{d \Lambda_1}{dt} & = & -\gamma_{\perp} [(1+i \Delta) \Lambda_1+n_1 \psi_1] \nonumber \\
 \frac{dn_1}{dt} & = & -\gamma_{\parallel}\left[  n_1-1-\frac{1}{2} \left( \psi_1^* \Lambda_1+\psi_1 \Lambda_1^* \right)  \right]
\end{eqnarray}
In their present form, Eqs.(A7) and (A8) represent a system of coupled differential-delayed equations that should be integrated with the initial conditions $n_1(0)$, $\Lambda_1(0)$ and $\psi_1(t)$ for $0<t<T_R$.  The common mean-field and single-mode laser model \cite{L1,L2} is obtained from Eqs.(A7) and (A8) by taking the further limits $T=1-R \rightarrow 0$, $T_a \rightarrow 1$,  with $C \equiv gl/T$ finite, and assuming that the injected and intracavity fields $f_1(t)$ and $\psi_1(t)$ vary slowly with respect to time over one cavity transit time $T_R$. Under such assumptions, the delayed equation (A7) can be replaced by the following differential equation
\begin{equation}
\psi_1(t)+T_R \frac{d \psi_1}{dt} \simeq \sqrt{T} f_1(t) +\sqrt{T_a} \psi_1(t)-\frac{T}{2} \psi_1(t)-gl\Lambda_1(t)
\end{equation}
i.e.
\begin{equation}
 \frac{d \psi_1}{dt} \simeq -\kappa [\psi_1(t)+2C  \Lambda_1(t)]-\kappa_1 \psi_1(t)+ \kappa \frac{2 f_1(t)}{\sqrt{T}}
\end{equation}
where $\kappa$ and $\kappa_1$ are defined by Eqs.(16) and (19) given in the text. The system of coupled differential equations (A8) and (A10) is precisely the modified single-mode Lorenz-Haken laser model (18) given in the text. Note that the usual Lorenz-Haken laser model (15) is simply obtained after setting $f_1=0$ and $T_a=1$ (i.e. $\kappa_1=0$).


\begin{thebibliography}{}

\bibitem{L1}
See, for instance: H. Haken, {\it Light. Vol.II: Laser Light Dynamics} (North-Holland, Amsterdam, 1985). 

\bibitem{L2}
N. B. Abrahm, P. Mandel, and L.M. Narducci, Prog. Opt. {\bf 25}, 1
(1988).

\bibitem{L3}
A. E. Siegman, {\it Lasers} (University Science Books, Mill Valley, CA, 1986).

\bibitem{L4}
O. Svelto, {\it Principles of Lasers}, fourth ed.  (Springer, Berlin, 1998).

\bibitem{Stone1}
Y. D. Chong, Li Ge, Hui Cao, and A. D. Stone, Phys. Rev. Lett. {\bf
105}, 053901 (2010). See also: S. Longhi, Physics {\bf 3}, 61 (2010); C.F. Gmachl,  Nature {\bf 467},  37 (2010); A.D. Stone, Phys. Today {\bf 64}, 68 (2011); A. Lagendijk, Nature Photon. {\bf 5}, 252 (2011).

\bibitem{CC}
See, for instance: M. Cai, O. Painter and K. J. Vahala, Phys. Rev. Lett. {\bf 85}, 74 (2000).

\bibitem{LonghiPRA2010}
S. Longhi, Phys. Rev. A {\bf 82}, 031801(R) (2010).

\bibitem{StoneScience2011}
W. Wan, Y. Chong, Li Ge, H. Noh, A.D. Stone, and H. Cao, Science
{\bf 331}, 889 (2011).

\bibitem{StonePRL2011}
Y.D. Chong, Li Ge, and A.D. Stone, Phys. Rev. Lett. {\bf 106},
093902 (2011).

\bibitem{uffa}
Y. D. Chong and A. D. Stone, Phys. Rev. Lett. {\bf 107}, 163901 (2011).

\bibitem{uffa2}
S. Longhi, Phys. Rev. Lett. {\bf 107}, 033901 (2011).

\bibitem{uffa3}
S. Longhi, Phys. Rev. A {\bf 83}, 055804 (2011).

\bibitem{uffa3bis}
J. Yoon, K. Hee Seol, S. Ho Song, and R. Magnusson, 
Opt. Express {\bf 18}, 25702 (2010); J. W. Yoon, W.J. Park, K.J. Lee, S.H. Song,
and R. Magnusson, Opt. Express {\bf 19}, 20673 (2011).

\bibitem{StoneP}
H. Noh, Y. Chong, A.D. Stone, and H. Cao, arXiv:1110.4950

\bibitem{uffa4}
M. Pu, Q. Feng, M. Wang, C. Hu, C. Huang, X. Ma, Z. Zhao, C. Wang, and X, Luo, Opt. Express {\bf 20}, 2246 (2012). 

\bibitem{uffa5}
Shourya Dutta-Gupta, O.J.F. Martin, S. Dutta Gupta, and G. S. Agarwal, Opt. Express {\bf 20}, 1330 (2012).

\bibitem{Newell}
A.C. Newell, J.V. Moloney, {\it Nonlinear Optics} (Addison- Wesley, Redwood City, California, 1991), Section 5.c.

\bibitem{Haus}
H. A. Haus, IEEE J. Sel. Top. Quantum Electron. {\bf 6}, 1173 (2000).

\bibitem{chaos}
H. Haken, Phys. Lett. A {\bf 53}, 77 (1975); C.O. Weiss, N.B. Abraham, and U. H\"{u}bner, Phys. Rev. Lett. {\bf 61}, 1587 (1988); R. Roy, T.W. Murphy, T.D. Maier, Z. Gills, and E.R. Hunt, Phys. Rev. Lett. {\bf 68}, 1259 (1992).

\bibitem{note}
We note that, as the field $E(t)$ emitted by the laser system $\mathbf{S}$ and the field $f_1(t)$ injected into the CPA system $\mathbf{S}'$ are outgoing and ingoing fields, respectively, Eq.(7) means that the CPA device perfectly absorbs the "mirror-reflected" signal $E(t-z/c)$ of the output laser signal $E(t+z/c)$, attenuated by $\sqrt{R}$.

\bibitem{note2}
An an example, let us consider values of $\kappa / \gamma_{\perp}$, $\gamma_{\parallel} / \gamma_{\perp}$ and $2C$ where the Lorenz-Haken instability does not sets in. In this case and assuming $\Delta=0$, the stationary field solution to Eqs.(15), given by $\psi=\sqrt{2C-1} \exp(i \varphi)$, is stable  ($\varphi$ is an arbitrary phase). Such a solution  corresponds to the laser system $\mathbf{S}$ emitting a continuous wave. Correspondingly, it can be readily shown that the system $\mathbf{S}'$ with the injected signal $\psi$ admits of a single stationary (and stable) solution, $\psi_{1,1}=\psi$,  for $2C<25$, corresponding to the CPA condition. However, for $2C>25$, there appear two other solutions, given by $\psi_{1,2}=-(1/6) (\sqrt{2C-1}) \exp(i \varphi)+(1/6) (\sqrt{2C-25}) \exp(\i \varphi)$,  and $\psi_{1,3}=-(1/6) (\sqrt{2C-1}) \exp(i \varphi)-(1/6) (\sqrt{2C-25}) \exp(\i \varphi)$. As the former solution is linearly unstable, the latter one is stable and can be a second attractor to Eqs.(20), which does not correspond to a CPA regime. However, large perturbations in the system can switch the dynamics toward the CPA basin of attraction. 

\bibitem{FM1}
S. E. Harris and R. Targ, Appl. Phys. Lett. {\bf 5}, 202  (1964); A. Yariv, J. Appl. Phys. {\bf 36}, 388  (1965); S. E. Harris and O. P. McDuff, IEEE J. Quantum Electron. {\bf QE-1}, 245
(1965).

\bibitem{FM2}
S. Longhi and P. Laporta, Appl. Phys. Lett. {\bf 73}, 720 (1998); S. Longhi and P. Laporta, Phys. Rev. A {\bf 60}, 4016 (1999).

\bibitem{ufff}
S. Longhi, Phys. Rev. E {\bf 63}, 037201 (2001).

\bibitem{FM3}
S. Longhi, S. Taccheo, and P. Laporta, Opt. Lett. {\bf 22}, 1642 (1998).

\bibitem{FM4}
In the transition region between the FM-operation and the  FM-mode locking regimes, the laser exhibits a strongly
enhanced sensitivity to external noise, which includes large transient energy amplification of perturbations in
the deterministic case and enhancement of field fluctuations in presence of a continuous stochastic noise [see: S. Longhi and P. Laporta, Phys. Rev. E {\bf 61}, R989 (2000)].


\end{thebibliography}
\end{document}